\documentclass[12pt]{article}

\newcommand{\munu}{^\mu_{\phantom{\mu}\nu}}

\begin{document}
\begin{center}
{\large{\bf  Einstein-Podolsky-Rosen correlation
seen from moving observers}}
\vskip .5 cm
{\bf Hiroaki Terashima and Masahito Ueda}
\vskip .4 cm
{\it Department of Physics, Tokyo Institute of Technology\\
Tokyo 152-8551, Japan}
\vskip 0.5 cm
\end{center}

\begin{abstract}
Within the framework of relativistic quantum theory,
we consider the Einstein-Podolsky-Rosen (EPR)
gedanken-experiment in which measurements of the spin
are performed by moving observers.
We find that the perfect anti-correlation
in the same direction
between the EPR pair no longer holds
in the observers' frame.
This does not imply a breakdown of
the non-local correlation.
We explicitly show that
the observers must measure the spin in
appropriately chosen \emph{different} directions
in order to observe the perfect anti-correlation.
This fact should be taken into account in
utilizing the entangled state
in quantum communication by moving observers.
\end{abstract}

In 1935, Einstein, Podolsky, and Rosen
(EPR)~\cite{EiPoRo35} proposed
a gedankenexperiment in an attempt
to show that the description of physical reality
by quantum theory is not complete.
A variant EPR gedankenexperiment was
put forth by Bohm~\cite{Bohm89}
in which a pair of spin-$1/2$ particles
with total spin zero are moving
in opposite directions.
This state has a remarkable property,
known as the EPR correlation,
that measurements of the spin show perfect
anti-correlation in whichever direction
and at however remote places they are performed.
In this Letter, we formulate Bohm's version of
the EPR correlation within the framework of
relativistic quantum theory
and consider a situation in which
measurements are performed by moving observers.
We here focus on the role played by the moving observers
from the viewpoint of the unitary transformation of the spin
under the Lorentz transformation~\cite{Weinbe95}.
We find that the perfect anti-correlation
in the \emph{same} direction between the spins of an EPR pair
deteriorates in the moving observers' frame
in the following sense:
even if the observers measure the spins
in the same direction in their frame
(which is also the same in the laboratory frame),
the results of measurements are not always anti-correlated.
That is, the perfect anti-correlation
in the \emph{same} direction is not Lorentz invariant.
The perfect entanglement should, however, be preserved
since the Lorentz transformation is
a local unitary operation~\cite{AlsMil02}.
The perfect anti-correlation is thus maintained
in \emph{different} directions.
We then show
that the degree of the violation of
Bell's inequality~\cite{Bell64,CHSH69} decreases
with increasing the velocity of the observers.
This also is a consequence of a local unitary operator
associated with the Lorentz transformation
which does not imply a breakdown of non-local correlations.
Our aim is to explore effects of the relative motion
between the sender and receiver in quantum communication.

Consider a massive particle with mass $M$.
In the rest frame of the particle,
the four-momentum is given by
the rest momentum $k^\mu=(Mc,0,0,0)$.
In this frame,
the state $|k,\sigma\rangle$ is specified
in terms of the eigenvalues of the Hamiltonian $H$,
the momentum operator $\vec{P}$, and
the $z$-component of
the total angular momentum operator $\vec{J}$ as
$H |k,\sigma\rangle=Mc^2 |k,\sigma\rangle$,
$\vec{P}|k,\sigma\rangle = 0$, and
$J^3 |k,\sigma\rangle=\sigma\hbar |k,\sigma\rangle$,
respectively.
Since the momentum $k^\mu$ is invariant under
the spatial rotation group $SO(3)$,
a rotation $R\munu$ is represented by
a $(2j+1)$-dimensional unitary matrix $D^{(j)}(R)$,
\begin{equation}
U(R)\, |k,\sigma\rangle  = \sum_{\sigma'}
    D_{\sigma'\sigma}^{(j)}(R)\,|k,\sigma'\rangle,
\label{eq:rotj}
\end{equation}
where $j$ is an integer or a half-integer
and $-j\le\sigma\le j$.
Note that $j$ is the spin of the particle and
$\sigma$ its $z$-component
because the orbital angular momentum is absent
in the rest frame of the particle.

In the laboratory frame, the four-momentum of the particle
has a generic form
$p^\mu=(\sqrt{|\vec{p}|^2+M^2c^2},p^1,p^2,p^3)$,
which is obtained by performing
a standard Lorentz transformation $L(p)\munu$
on the rest momentum $k^\mu$,
i.e. $p^\mu=L(p)\munu k^{\nu}$, where $\mu,\nu=0,1,2,3$ and
repeated indices are assumed to be summed.
An explicit form of $L(p)\munu$ is written as
\begin{eqnarray}
L(p)^0_{\phantom{0}0} &=& \gamma, \qquad
L(p)^0_{\phantom{0}i}=L(p)^i_{\phantom{i}0}=p^i/Mc,
\nonumber \\
L(p)^i_{\phantom{i}k} &=& \delta_{ik}+
(\gamma-1)\,p^i\,p^k/|\vec{p}|^2,
\label{eq:explp}
\end{eqnarray}
where $\gamma=\sqrt{|\vec{p}|^2+M^2c^2}/Mc$ and $i,k=1,2,3$.
Using the unitary operator corresponding to
$L(p)\munu$, the state in this frame is defined by
$|p,\sigma\rangle\equiv\,U(L(p))\, |k,\sigma\rangle$.

Consider now a situation
in which an observer is moving
with respect to the laboratory frame,
and let $\Lambda\munu$ be the corresponding
Lorentz transformation.
For this observer, the state of the particle
is described by
$U(\Lambda)\, |p,\sigma\rangle
=U(L(\Lambda p))\,U(W(\Lambda,p))\, |k,\sigma\rangle$,
where
\begin{equation}
W(\Lambda,p)\munu=
\left[L^{-1}(\Lambda p)\,\Lambda\,L(p)\right]\munu
\label{eq:defw}
\end{equation}
is the Wigner rotation~\cite{Wigner39}.
The Wigner rotation is an element of
the spatial rotation group $SO(3)$ since
it leaves the rest momentum $k^\mu$ unchanged
by the definition of $L(p)\munu$.
It follows then from Eq.~(\ref{eq:rotj}) that
\begin{equation}
 U(\Lambda)\, |p,\sigma\rangle
  = \sum_{\sigma'} D_{\sigma'\sigma}^{(j)}(W(\Lambda,p))
    \,|\Lambda p,\sigma'\rangle.
\label{eq:trans}
\end{equation}

To be specific,
suppose that a massive spin-$1/2$ particle
(e.g. electron) is moving along the $x$-axis with
the laboratory-frame four-momentum given by
$p^\mu=(Mc\,\cosh\xi,Mc\,\sinh\xi,0,0)$,
where the rapidity $\xi$ is related to
the velocity of the particle $v$ by $v/c=\tanh\xi$.
In this case,
the Lorentz transformation (\ref{eq:explp}) becomes
\begin{equation}
L(p)\munu=
 \left(\begin{array}{cccc}
    \cosh\xi &  \sinh\xi & 0 & 0 \\
    \sinh\xi &  \cosh\xi & 0 & 0 \\
           0 &         0 & 1 & 0 \\
           0 &         0 & 0 & 1
  \end{array}\right).
\end{equation}
Suppose that an observer is moving along the $z$-axis
with the velocity given by $V=c\,\tanh\chi$.
The corresponding Lorentz transformation reads
\begin{equation}
\Lambda\munu=
   \left(\begin{array}{cccc}
   \cosh\chi & 0 & 0 & -\sinh\chi \\
           0 & 1 & 0 & 0          \\
           0 & 0 & 1 & 0          \\
  -\sinh\chi & 0 & 0 & \cosh\chi
   \end{array}\right).
\end{equation}
The Wigner rotation (\ref{eq:defw}) is then
reduced to a rotation about the $y$-axis.
The angle $\delta$ of this rotation is given by
\begin{equation}
\tan\delta = \frac{\sinh\xi\sinh\chi}{\cosh\xi+\cosh\chi}. 
\label{eq:del}
\end{equation}
This angle $\delta$ becomes $\xi\chi/2$
in the limit of $\xi\to0$ and $\chi\to0$,
and $\pi/2$ in the limit of $\xi\to\infty$ and $\chi\to\infty$.
If either $\xi=0$ or $\chi=0$, $\delta$ vanishes.
For the case of the spin-$1/2$ particle,
rotations are represented by the Pauli matrices.
Using the Pauli matrix $\sigma_y$,
the transformation law (\ref{eq:trans}) thus becomes
\begin{eqnarray}
U(\Lambda)\,|p,\uparrow\rangle &=&
 \cos\frac{\delta}{2}\, |\Lambda p,\uparrow\rangle
 +\sin\frac{\delta}{2}\, |\Lambda p,\downarrow\rangle, \label{eq:tr1}\\
U(\Lambda)\,|p,\downarrow\rangle &=&
 -\sin\frac{\delta}{2}\, |\Lambda p,\uparrow\rangle
 +\cos\frac{\delta}{2}\, |\Lambda p,\downarrow\rangle, \label{eq:tr2}
\end{eqnarray}
where $\uparrow=+1/2$ and $\downarrow=-1/2$.
That is, the spin is rotated about the $y$-axis through
the angle $\delta$ in the observer's frame.

A physical picture of this spin rotation is as follows.
The Lorentz transformation $\Lambda\munu$
``rotates'' the direction of the momentum
from $p^\mu$ to $\Lambda p^\mu$.
The spin is simultaneously rotated
by this Lorentz transformation
since the spin is coupled with the momentum
in relativistic quantum theory.
In non-relativistic quantum theory,
the Galilean transformation rotates
the direction of the momentum but not the spin.
Note that the angle of rotation for the spin
is \emph{not} equal to that for the momentum
because $\Lambda\munu$ is not a spatial rotation
but a Lorentz transformation.
The rotation of the spin comes from the fact that
$\Lambda L(p)\munu$ and $L(\Lambda p)\munu$
are not equal even though both of them
bring the momentum $k^\mu$ to $\Lambda p^\mu$.

We use the transformation law obtained above
to analyze the relativistic EPR correlation.
Suppose that a pair of spin-$1/2$ particles
with total spin zero are moving away
from each other in the $x$ direction.
This situation is described by the state
\begin{equation}
 |\psi\rangle=\frac{1}{\sqrt{2}}
   \Bigl[\, |p_+,\uparrow\rangle|p_-,\downarrow\rangle-
      |p_+,\downarrow\rangle|p_-,\uparrow\rangle \,\Bigr],
\label{eq:relepr}
\end{equation}
where
$p^\mu_\pm=(Mc\,\cosh\xi,\pm Mc\,\sinh\xi,0,0)$.
Unlike the non-relativistic case,
we need to explicitly specify
the motion of the particles
because the Lorentz transformation of the spin
depends on the momentum as in Eq.~(\ref{eq:trans}).
In the EPR experiment,
we have two observers who perform measurements
on the particles, respectively.
Here we assume that both observers are moving
in the $z$ direction at the same velocity $V$.
Using the transformation formulas
(\ref{eq:tr1}) and (\ref{eq:tr2}), we find that
the moving observers see the state (\ref{eq:relepr}) as
\begin{eqnarray}
 U(\Lambda)|\psi\rangle &=&
  \frac{1}{\sqrt{2}}
 \Biggl[\cos\delta\Bigl(
 |\Lambda p_+,\uparrow\rangle|\Lambda p_-,\downarrow\rangle-
 |\Lambda p_+,\downarrow\rangle|\Lambda p_-,\uparrow\rangle\Bigr)
   \nonumber \\
&& \qquad+\sin\delta\Bigl(
    |\Lambda p_+,\uparrow\rangle|\Lambda p_-,\uparrow\rangle+
    |\Lambda p_+,\downarrow\rangle|\Lambda p_-,\downarrow\rangle
    \Bigr) \Biggr],
\label{eq:relstate}
\end{eqnarray}
where $\delta$ is given by Eq.~(\ref{eq:del}).
Note that the spins of the two particles
are rotated in opposite directions
because they are moving oppositely.

From Eq.~(\ref{eq:relstate}), we find that
the measurements of the spin $z$-component
will no longer show perfect anti-correlation.
Note that the two observers are at rest in the common frame,
since they are moving in the same direction at the same velocity
with respect to the laboratory frame.
The directions that are the same in this observers' frame are
also the same in the laboratory frame.
Thus, in non-relativistic theory,
the measurements in the same direction
of the observers' frame must be perfectly anti-correlated.
Moreover, the $z$ direction in the observers' frame
is identical to that in the laboratory frame
since the observers are moving along the $z$-axis.
Nevertheless, the anti-correlation in the $z$ direction
is reduced in relativistic theory.
(On the other hand, the measurements of
the spin $y$-component are perfectly anti-correlated
for any $\xi$ and $\chi$.)
Note that in Eq.~(\ref{eq:relstate})
the spin-singlet state is mixed with
the spin-triplet state.
This is because
while the spin-singlet state is invariant
under spatial rotations, it is not invariant under
Lorentz transformations.
That is, the Poincar\'{e} group $ISO(1,3)$
is larger than the spatial rotation group $SO(3)$.

Let us now examine Bell's inequality
in the same situation~\cite{Czacho97}.
Let $Q$ and $R$ be operators
on the first particle corresponding to
the spin $z$- and $y$- components, respectively.
Similarly, let $S$ and $T$ be operators
on the second particle corresponding to
the spin component in the directions
$(0,-1/\sqrt{2},-1/\sqrt{2})$ and
$(0,-1/\sqrt{2},1/\sqrt{2})$, respectively.
Then, for the state (\ref{eq:relstate}),
we obtain
\begin{equation}
\langle QS\rangle+\langle RS\rangle+
\langle RT\rangle-\langle QT\rangle= 2\sqrt{2}\cos^2\delta.
\end{equation}
The right-hand side
decreases with increasing the velocity of the observers
and with increasing that of the particles,
and vanishes in the limit of
$\xi\to\infty$ and $\chi\to\infty$.

Does this result imply
a breakdown of the EPR correlation or
a revival of the local realism?
The answer is, of course, no.
If the directions of measurements are rotated
about the $y$-axis through $\delta$ for the first particle
and through $-\delta$ for the second
in accordance with the spin rotation,
the measurements of the spin are perfectly anti-correlated
and Bell's inequality remains maximally violated.
While the perfect anti-correlation
in the \emph{same} direction no longer holds,
the perfect anti-correlation is maintained
in the appropriately chosen different directions.
Naive measurements lead to wrong conclusions.

In conclusion, in the relativistic EPR experiment
with a pair of massive spin-$1/2$ particles,
the spin-singlet state is mixed with the spin-triplet state
if the measurements are performed
by orthogonally moving observers.
This is because the Poincar\'{e} group
is larger than the spatial rotation group.
Therefore, the perfect anti-correlation in the same
direction deteriorates.
We must carefully choose the directions of measurements
to obtain the perfect anti-correlation and
the maximal violation of Bell's inequality
which are utilized
in quantum communication~\cite{Ekert91,BeBrMe92,BBCJPW93}.
We can also obtain a similar conclusion
in the case of massless particles.

After this manuscript was prepared,
the authors become aware of the work by
Alsing and Milburn~\cite{AlsMil02}.
Although they have considered a similar situation,
there are essential differences from ours.
They have discussed the Lorentz invariance of
entanglement using the spin-triplet state.
Here we have discussed the change of
the anti-correlation due to the Lorentz transformation,
using the spin-singlet state.
Note that the entanglement is independent of the basis
but the correlation depends on the basis to measure,
and the spin-triplet state does not have
the property of the anti-correlation
in all the directions,
unlike the spin-singlet state.

\section*{Acknowledgements}
H.T. was partially supported
by JSPS Research Fellowships for Young Scientists.
This work was supported by a Grant-in-Aid for
Scientific Research (Grant No. 11216204)
by the Ministry of Education, Science, Sports, and Culture
of Japan, by the Toray Science Foundation,
and by the Yamada Science Foundation.

\end{document}